\documentclass[final,5p,times,twocolumn]{elsarticle}

\usepackage{color}
\usepackage{amsmath}

\biboptions{numbers,sort&compress}

\journal{Physica E}

\newcommand{\e}{{\rm e}}
\newcommand{\rmd}{{\rm d}}

\newcommand{\eps}{\epsilon}

\newcommand{\kB}{k_{\rm B}}

\definecolor{DarkGreen}{rgb}{0,0.7,0}

\begin{document}

\begin{frontmatter}

\title{Thermoelectricity without absorbing 
energy from the heat sources}

\author{Robert.~S.~Whitney$^1$, Rafael~S\'anchez$^2$, Federica~Haupt$^3$, and Janine~Splettstoesser$^4$}
\address{
$^1$ Laboratoire de Physique et Mod\'elisation des Milieux Condens\'es (UMR 5493), 
Universit\'e Grenoble Alpes and CNRS, \\ 25 rue des Martyrs, B.P.\ 166, 38042 Grenoble, France\\
 $^2$ Instituto de Ciencia de Materiales de Madrid, CSIC, Cantoblanco, 28049 Madrid, Spain \\
$^3$ JARA Institute for Quantum Information, RWTH Aachen University, D-52056 Aachen, Germany\\
$^4$ Department of Microtechnology and Nanoscience (MC2), Chalmers University of Technology, 
SE-41296 G\"oteborg, Sweden
\\
\vskip -8mm}

\begin{abstract}
We analyze the power output of a quantum dot machine coupled to two electronic reservoirs via thermoelectric contacts, and to 
two thermal reservoirs -- one hot and one cold. This machine is a nanoscale analogue of a conventional thermocouple heat-engine, in which the \textit{active} region being heated is unavoidably also exchanging heat with its cold environment.
Heat exchange between the dot and the thermal reservoirs is treated as a capacitive coupling to electronic fluctuations in localized levels, modeled as two additional quantum dots.  The resulting multiple-dot setup is described using a master equation approach.
We observe an ``exotic'' power generation, which remains finite even when the heat absorbed from the thermal reservoirs is zero (in other words the heat coming from the hot reservoir all escapes into the cold environment).
This effect can be understood in terms of a non-local effect in which the heat flow from heat source to the cold environment generates power via a mechanism which we refer to as \textit{Coulomb heat drag}.
It relies on the fact that there is no relaxation in the quantum dot system, 
so electrons within it have a non-thermal energy distribution.
More poetically, one can say that we find a spatial separation of the first-law of thermodynamics (heat to work conversion) from the second-law of thermodynamics (generation of entropy).
We present circumstances in which this \textit{non-thermal} system can generate more power than any conventional macroscopic thermocouple (with local thermalization), even when the latter works with Carnot efficiency.  

\vskip 4mm
\hrule
\vskip 4mm

\hskip -3.5mm {\bf Highlights}

\vskip -4.1mm
\hskip 25mm $\bullet\, $  Model of a quantum-dot energy-harvester: a quantum analogue of a thermocouple. 

\hskip 25mm $\bullet\, $  Two thermal reservoirs (heat source and cold environment) and two electric contacts.

\hskip 25mm $\bullet\, $  Non-local Coulomb heat drag effects, non-locality of laws of thermodynamics.

\hskip 25mm $\bullet\, $  Finite power output without absorbing heat from the thermal reservoirs.

\hskip 25mm $\bullet\, $  Can beat a Carnot-efficient conventional thermocouple under equivalent conditions.

\vskip2mm
\end{abstract}

\begin{keyword}
quantum thermodynamics \sep thermocouples \sep thermoelectricity \sep quantum transport 
\sep energy harvesting \sep Coulomb drag



\end{keyword}

\end{frontmatter}

\begin{figure}[t!]
\centerline{\includegraphics[width=0.7\columnwidth]{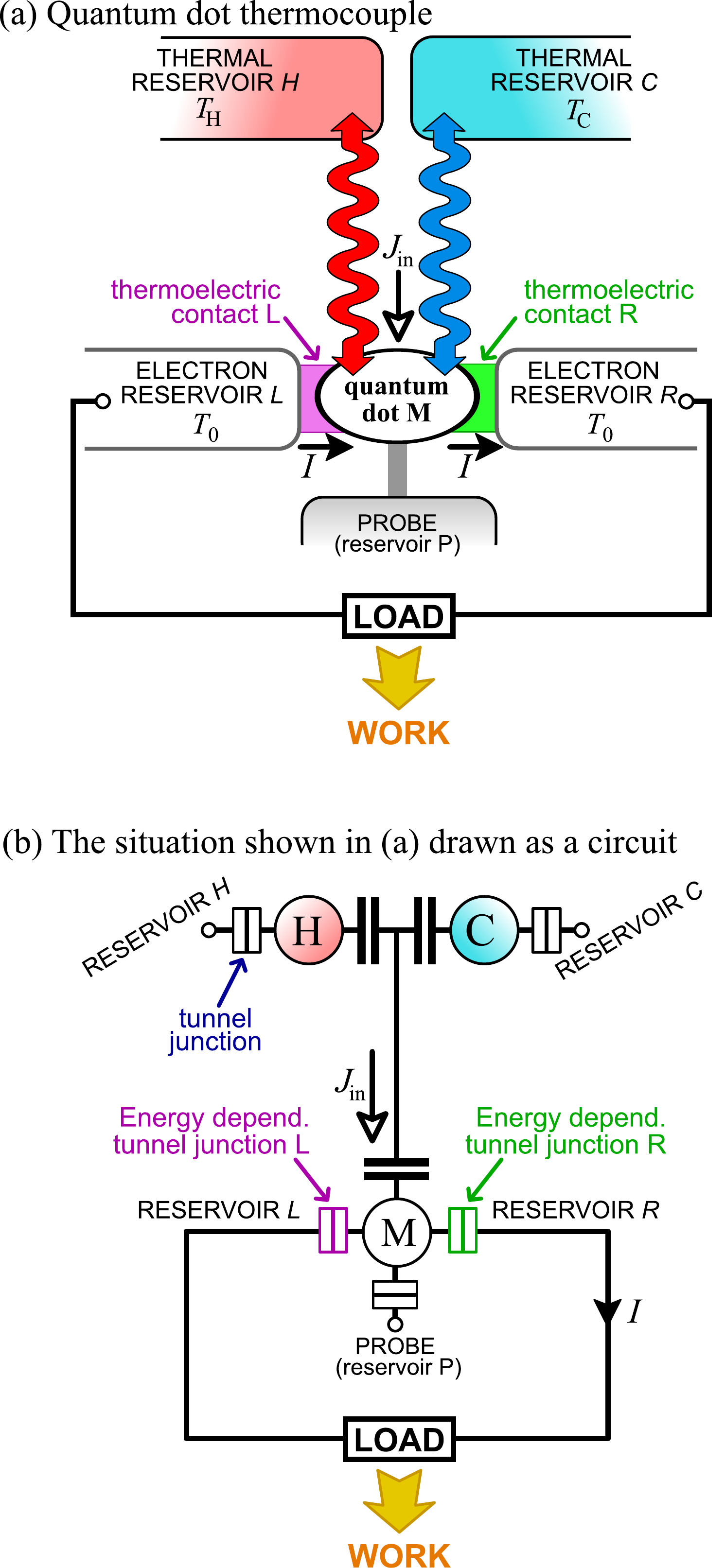}}
\caption{\label{Fig:system}
(a) A sketch of a quantum dot thermocouple in which 
the electrons in the central part of the thermocouple (the quantum dot) exchange heat 
with its cold environment (Reservoir C)  as well as the heat source (Reservoir H). 
The couplings of the dot to 
reservoirs L and R have an energy dependence chosen to ensure that each one has the opposite thermoelectric response.   As a result they form a thermocouple, which can be used to generate electrical power.   The load is taken to be a device that converts the electrical power into some other form of useful work
 (it could be a motor converting the electrical work into mechanical work).
(b) A circuit picture of the set-up in (a), in which the exchange of energy with the heat source and the environment are modeled by a capacitive coupling to electronic fluctuations in those reservoirs, which we take to be thermally activated hopping of electrons between the bulk and a localized state, indicated by the two upper circles (H and C).  Thus each of the three circles represents a two-state system,
with occupations 0 or 1.
}
\end{figure}

{\it Dedicated to Markus B\"uttiker: In addition to his human qualities, we remember Markus fondly for the inspiring discussions we had.
We think that this work would have led to another lively and enjoyable debate.}
\section{Introduction}

There is great current interest in quantum and nanoscale systems that convert heat into electric current~\cite{Giazotto06,Casati-review,Sothmann15}.
The simplest such systems are those that exhibit a thermoelectric effect
\cite{Shakouri11,Cahill14}: they do the conversion 
in a steady-state (DC) manner, 
and so avoid the need for pumping cycles relying on time-dependent couplings.
Quantum dots are particularly promising in this respect, and various applications have been proposed  and (at least partially) realized experimentally, including thermoelectric engines~\cite{Humphrey02,Esposito09-10,Sanchez11,Ruokola12,Sothmann12,Sothmann12b,Jordan13,DSanchez13,Thierschmann13,Jiang2014b,Whitney14,Whitney15,Roche15,Hartmann15,Thierschmann15b}, refrigerators~\cite{Edwards93,Arrachea07,Rey07,Prance09,Venturelli13}, thermal rectifiers~\cite{Ruokola11,Jiang2015}, and hybrid refrigerator power-sources \cite{Entin15}.  
In addition, the simplicity of these thermoelectric quantum dots
makes them ideal model systems for the study of quantum thermodynamics. 

The most developed theory of non-equilibrium thermodynamics, known as irreversible thermodynamics
\cite{book:irreversible-thermodyn}, assumes that the shortest lengthscale in the system is that on which the particles thermalize (the inelastic scattering length).  Then, any system coupled between two reservoirs at very different temperatures has a well-defined local temperature everywhere within it.  
However, as we reduce the size of the circuit elements into the nanoscale regime, this ceases to be the case.  A system that is much smaller than the electron thermalization length will be in a {\it non-thermal state} whenever it is coupled to reservoirs that are at significantly different temperatures or electrochemical potentials.
In other words, the distribution of the excitations in the system 
cannot be described in terms of a thermal distribution.

In general, we wish to answer the question: what new physics can emerge in a quantum system operating in this non-thermal regime?
This work presents a first response to this question, 
in the context of the thermoelectric quantum dot system sketched 
in Fig.~\ref{Fig:system}(a).  This can be considered as a miniaturized version of the usual macroscopic thermocouple
power generator, in which the macroscopic metal reservoir between the two thermoelectrics has been replaced by the quantum dot M. 
Our system is similar to three-terminal energy harvesters considered in Refs.~\cite{Sanchez11,Sanchez13}, which 
 separate the conductor from a heats source with which it exchanges energy but no particles.
See also Refs.~\cite{Rutten09,Entin10,Ruokola12,Jiang2012}, which
consider related models with bosonic heat sources.
However, instead of the dot being coupled only to the heat source (reservoir H), we consider the case when
it is also coupled to the  cold environment (reservoir C).  Since the environment coupling is never negligible,
we argue that this is the generic case.  It is certainly the case in conventional thermocouples, 
where it is well known that the region between the two thermoelectrics
is not as hot as the heat-source, because it also exchanges heat with the cold environment. 
With this in mind,  
the simplest case would be when the temperature of the cold environment ${\rm C}$ is the same as the temperature $T_{0}$ of the electrical circuit in which the thermocouple is inserted (reservoirs L and R). 
However, we consider the more general case where $T_\text{C}$ is different from $T_0$, for example due to Joule heating in the wires or load ($T_0 > T_\text{C}$).  
Following what 
is known about conventional thermocouples, one would expect the energy available to produce
electricity to  be equal to the heat that flows in from reservoir H minus the heat that is lost as it flows out
into reservoir C.  We call 
this quantity $J_\text{in}$, and study how the power generated by a 
quantum dot thermocouple, $P_{\rm gen}$, depends on it.     
 
In this work, we show that the quantum-dot thermocouple can generate power even when the
total heat absorbed from the thermal reservoirs (H and C) is zero ($J_\text{in}=0$).  
Instead, the dot extracts heat from the electronic reservoirs (L and R) and converts it into electrical power. 
It might appear that this ``exotic'' power generation violates the laws of thermodynamics
(since the thermal reservoirs provide no energy to allow the dot to convert heat into work), 
but we show that this is not the case. 
It can be explained in terms of a non-local heat drag effect, in which the heat flow from H to C can induce heat and charge currents in the circuit 
(i.e. through the thermoelectric contacts to reservoirs R and L) even when $J_\text{in}=0$. 
We show that this exotic power generation only arises because no thermalization occurs within the quantum dot,
meaning that the dot is maintained in a non-thermal state by its contact with the hot and cold reservoirs.
As such it has no analogue in conventional thermocouples.  We argue that as a result a quantum-dot thermocouple can achieve power outputs
larger than any conventional thermocouple  
(even one working with Carnot efficiency) over a broad range of parameters. This is not because our device has an efficiency higher than Carnot efficiency
(this is forbidden by the laws of thermodynamics), but because it can also extract useful work from the non-local drag effect. This is something a conventional thermocouple cannot do, irrespective of how efficient it is.

This work is organized as follows.  
In Sec.~\ref{Sect:quantum-dot}, we introduce our model of the quantum dot thermocouple. Sec.~\ref{Sect:V=0} discusses the basic mechanisms behind  the rectification of heat fluctuations in our device, in the simpler case where it generates no power.
Sec.~\ref{Sect:finite-power} gives results for the power generation, which Sec.~\ref{Sect:interpretation} 
explains in terms of non-local heat drag, shows that it obeys the laws of thermodynamics, 
and discusses potential experimental implementation.  
Sec.~\ref{Sect:thermalize} shows that the effect is suppressed if relaxation processes cause the state of dot M to become thermal.
Finally, in Sec.~\ref{Sect:thermocouple}, we compare the power output of the quantum dot thermocouple with that of a conventional thermocouple for  a broad range of parameters. 

\section{Model of the quantum-dot thermocouple}
\label{Sect:quantum-dot}
 
The quantum dot M, shown in Fig.~\ref{Fig:system}(a), 
can exchange energy (but not charge) with a hot (H) and a cold (C) thermal reservoir. 
In the following we are interested in the resulting amount of electrical power that this quantum dot can produce by driving an electric current between the left (L) and right (R) electron reservoirs, to which it is tunnel-coupled. Importantly, while the reservoirs are characterized by temperatures, the quantum dot is  
generally in a \textit{non-thermal} state. This is a crucial point of our paper and an important difference to conventional macroscopic thermocouples, which are large enough such that every element of the thermocouple is able to thermalize. 

The energy exchange of the quantum dot M with the thermal reservoirs H and C occurs via a capacitative coupling to thermal fluctuations in these reservoirs.  
They are modeled as thermally activated hopping of electrons between the bulk of each of the two thermal reservoirs and a localized state on their surface.  The localized states can be thought of as quantum dots, which enables us to recast the problem as the triple-dot circuit shown in 
Fig.~\ref{Fig:system}(b).  Here, the localized states associated with reservoirs H and C are the upper red 
and blue dots.
Apart from thermoelectric purposes, similar models have been useful for relaxation times 
detection~\cite{Schulenborg14}, Maxwell demon physics~\cite{Strasberg13,Koski15}, and the operation of stochastic logic gates~\cite{Pfeffer15}. 

We therefore now consider the three capacitively coupled quantum dots M, H, and C, in contact with four electronic reservoirs H, C, L, and R, as shown in Fig.~\ref{Fig:system}(b). We assume that the on-site 
Coulomb repulsion is very large, such that doubly occupation of any dot can be neglected.  For simplicity, we neglect the spin degree of freedom of the electrons, which, in this case, does not alter the working principle of the device, see also~\cite{Sothmann15} (note however, that in general the spin might play an important role in quantum-dot based heat engines~\cite{Juergens13}). The isolated system of three capacitively coupled quantum dots can then be described by the model Hamiltonian
\begin{equation}
\mathcal{H} = \sum_{i=\text{M,H,C}}\epsilon_i \ \hat{n}_i+\sum_{i,j=\text{M,H,C}}U_{ij}\ \hat{n}_i\hat{n}_j \ ,
\label{eq:Hamiltonian}
\end{equation}
with the number operator $\hat{n}_i$ counting the number of electrons ($0$ or $1$) on each of the three dots, $i=\text{M,H,C}$. Here, $\epsilon_i$ are the energies for single occupation of the different dots and $U_{ij}$ represents the Coulomb charging energy that needs to be paid to have both dots $i$ and $j$ occupied simultaneously. They play an essential role, as they mediate the energy exchange between the different dots. The eigenstates of this Hamiltonian are given by the eight states $(m,h,c)$ with eigenenergies $E_{mhc}$, where the labels $m,h,c$ can take the values $0$ (empty) and $1$ (full) depending on the occupation of the respective dots M, H, and C.  

Transitions between eigenstates of the triple-dot system can occur due to the tunnel coupling of these dots to the electronic reservoirs. 
These reservoirs are kept at temperatures $T_\alpha$ and electrochemical potentials $\mu_\alpha$. Here we take the temperature of the hot reservoir  to be always larger than the temperature of the cold reservoir, $T_\text{H}>T_\text{C}$. The left and right reservoirs are assumed to be at the same temperature, $T_\text{L}=T_\text{R}=T_\text{0}$ for $T_0 < T_\text{H}$.
In this paper, we are interested in the power that the quantum-dot thermocouple can generate, by driving an electrical current against a potential difference (due to a load). A simple way of implementing this setting is the one where we take the electrochemical potential of the right  lead to be $\mu_\text{R}=eV>0$ (where $e<0$ is the electron charge) with respect to the equal electrochemical potentials of the other reservoirs, which we take as the reference energy here, $\mu_\text{H}=\mu_\text{C}=\mu_\text{L}\equiv0$. 

The coupling between the quantum dots and the reservoirs, as depicted in Fig.~\ref{Fig:system}(b), is characterized by the tunnel coupling strengths, $\gamma_\alpha$. We focus on a situation, where the dots are weakly coupled to the reservoirs, $\gamma_\alpha\ll\kB T_\alpha$. Importantly, since we are interested in the quantum dot M behaving as a thermoelectric, we require that $\gamma_\text{L}$ and $\gamma_\text{R}$ depend on energy~\cite{Sanchez11}.  Because of the capacitive coupling between the dots, this means in turn that they depend on the occupation $h$ and $c$ of the dots H and C, yielding $\gamma_\text{L}=\gamma_\text{L}(h,c)$ and $\gamma_\text{R}=\gamma_\text{R}(h,c)$. A strong energy-dependence of the tunnel-coupling can be achieved for example by coupling the quantum dot to the reservoirs via other resonant levels~\cite{Sanchez11,Bryllert02}. This is not required for the tunnel-coupling between the thermal baths and the nearby quantum dots. We therefore assume $\gamma_\text{H}$ and $\gamma_\text{C}$ to be energy-independent.

We are interested in the steady-state response of the system to different temperature in the heat baths, in the presence of 
a possible voltage drop across the reservoirs L and R, which results in power being generated across the load. 
In order to calculate the DC charge and heat currents, the occupation probabilities $p_{mhc}$ of the states $(m,h,c)$ of the triple-dot system are required. In the weak-coupling regime, these probabilities are given by the solution of the steady-state master equation~\cite{Beenakker92,book:Breuer-Petruccione},
\begin{equation}
0 = \sum_{m',h',c'} M_{mhc}^{m'h'c'} p_{m'h'c'}\ ,
\label{Eq:master}
\end{equation}
with $\sum_{mhc}p_{mhc}=1$. All non-diagonal elements of the transition matrix are determined by 
\begin{equation}
M_{mhc}^{m'h'c'} = \sum_{\alpha=\text{H,C,L,R}}\big[ \Gamma_\alpha \big]_{\, mhc}^{\, m'h'c'} \ , 
\label{Eq:M-in-terms-of-Gammas}
\end{equation}
namely by the sum over the rates of transition of the three-dot system from state $m'h'c'$ to state $mhc$ due to electron tunnelling into or out of reservoir $\alpha$, $\big[ \Gamma_\alpha \big]_{\, mhc}^{\, m'h'c'}$. The latter are given by
\begin{subequations}
\begin{align}
\Big[ \Gamma_\text{L} \Big]_{\, mhc}^{\, m'h'c'} \!\!
&=\delta_{m,1-m'}\delta_{h,h'}\delta_{c,c'}\ \gamma_\text{L} \big(h,c\big)f_\text{L}\left({\Delta_{mhc}^{m'h'c'} } \right) , \quad
\\
\Big[ \Gamma_\text{R} \Big]_{\, mhc}^{\, m'h'c'}  \!\!
&=\delta_{m,1-m'}\delta_{h,h'}\delta_{c,c'}\ \gamma_\text{R} \big(h,c\big) f_R\left({\Delta_{mhc}^{m'h'c'}{-}(m{-}m')e V} \right),\quad \ 
\\
\Big[ \Gamma_\text{H} \Big]_{\, mhc}^{\, m'h'c'}  \!\!
&=\delta_{m,m'}\delta_{h,1-h'}\delta_{c,c'}\ \gamma_\text{H} f_\text{H}\left({\Delta_{mhc}^{m'h'c'} } \right) ,
\\
\Big[ \Gamma_\text{C} \Big]_{\, mhc}^{\, m'h'c'}  \!\!
&= \delta_{m,m'}\delta_{h,h'}\delta_{c,1-c'}\ \gamma_\text{C}f_\text{C}\left({\Delta_{mhc}^{m'h'c'} } \right) ,
\end{align}
\end{subequations}
where $\delta_{n,n'}$ is a Kronecker delta, $\Delta_{mhc}^{m'h'c'}=E_{mhc}-E_{m'h'c'}$, and  $f_\alpha(x)= 1\big/\big(1+\e^{x/\kB T_\alpha}\big)$ is the Fermi function corresponding to lead $\alpha$.
The diagonal elements in Eq.~(\ref{Eq:master}) are directly found by using the condition that every column of the transition matrix sums to zero to fulfill probability conservation, $M_{mhc}^{mhc}=-\sum_{\left\{mhc\right\}\neq\left\{m'h'c'\right\}}M_{mhc}^{m'h'c'}$.

For any heat-engine, the relevant quantities are the generated power (a charge current flowing against a potential difference) and the heat flow into the device. We therefore consider the charge and heat currents into the dots from reservoir $\alpha$, 
\begin{subequations}
\label{Eq:currents}
\begin{align}
I_\alpha & =  e I^{ N}_\alpha
\label{Eq:Charge-current}\\
J_\alpha & =  I^{E}_\alpha - \mu_\alpha I^{N}_\alpha.
\label{Eq:Heat-current}
\end{align} 
\end{subequations}
Here the charge current is trivially related to the particle current, $I^{ N}_\alpha$, by multiplication with the electronic charge $e$, while the heat current is given by the rate with which energy changes, $I^{ E}_\alpha$, with respect to the energy that particles at the electrochemical potential would carry in the respective reservoir. 
Using the Clausius definition of entropy, we see that the rate of change of entropy in
reservoir $\alpha$ is connected to the heat current by
\begin{eqnarray}
\big( \rmd S_\alpha \big/ \rmd t\big)= -J_\alpha/T_\alpha .
\end{eqnarray}
The particle currents and energy currents~\footnote{
These currents are time-averaged, and do not include the noise that comes from short-time fluctuations. This noise has a negligible effect on the long-time (steady-state) 
conversion of heat into work, see for example Section III of Ref.~\cite{Whitney15}.
}  can be written using the transition rates $\Big[ \Gamma_\alpha \Big]_{\, mhc}^{\, m'h'c'}$ and the occupation probabilities $p_{mhc}$ of the state $(m,h,c)$ with the respective energy $E_{mhc}$ and occupation number $N_{mhc}=m+h+c$,
\begin{eqnarray}
I^{N}_\alpha &=& \sum_{mhc}\sum_{m'h'c'} (N_{mhc}-N_{m'h'c'})\ \Big[ \Gamma_\alpha \Big]_{\, mhc}^{\, m'h'c'} p_{m'h'c'}\ ,
\label{Eq:Particle-current}
\\
I^{E}_\alpha &=& \sum_{mhc}\sum_{m'h'c'} \big(E_{mhc}-E_{m'h'c'}\big)\ \Big[ \Gamma_\alpha \Big]_{\, mhc}^{\, m'h'c'} p_{m'h'c'}\ . 
\label{Eq:Energy-current}
\end{eqnarray}
The particle and energy currents in reservoir $\alpha$ are defined to be positive when they are directed flowing \textit{into} the triple-dot system. Consequently, the charge current in reservoir $\alpha$, Eq.~(\ref{Eq:Charge-current}), is positive when electrons flow into the reservoir, while the heat current in reservoir $\alpha$, Eq.~(\ref{Eq:Heat-current}), is positive when flowing into the triple-dot system.
Particle and energy currents are conserved, so summing $\alpha$ over all reservoirs gives 
\begin{equation}
0=\sum_\alpha I^{ N}_\alpha = \sum_\alpha I^{ E}_\alpha\ .
\label{Eq:conservation-of-currents}
\end{equation}
Charge current is conserved (so $I_\text{R}=-I_\text{L}$).  Heat current is not conserved since
$\sum_\alpha J_\alpha \ =\ - \mu_\text{R}I^{ N}_\text{R}$, which corresponds to the first law of thermodynamics. 
The power generated by the quantum-dot thermocouple is 
\begin{equation}
\label{Eq:power_out}
P_\text{gen}  \ =   - \mu_\text{R}I^{ N}_\text{R} \ = \ - V I_\text{R} \ .
\end{equation}
These quantities are evaluated using Eqs.~(\ref{Eq:currents}), and we will discuss them in detail in the following sections.
For now, we do not consider the probe reservoir P, shown in Fig.~\ref{Fig:system},
however we will discuss it in Sections \ref{Sect:thermalize} and \ref{Sect:thermocouple}.

\section{Results in the absence of a bias voltage}
\label{Sect:V=0}

We start by analysing the behaviour of the quantum dot thermocouple in Fig.~\ref{Fig:system} under the following two assumptions.
(i) The load's resistance is zero, so the bias $V$ remains zero when a current is flowing. 
(ii) The capacitive coupling between dot H and C is strong enough that $U_\text{HC}$ is very much bigger than all temperatures, so the probability that these two dots are occupied at the same time becomes vanishingly small. 
We will relax both these assumption in Section \ref{Sect:finite-power}.

For strong $U_\text{HC}$, the states (0,1,1) and (1,1,1) drop out of the problem. In this limit, the charge fluctuations that lead to transport through the conductor can be separated in two cycles. 
The first cycle is
\begin{subequations}
\label{Eq:cycles}
\begin{equation}
(0,0,0)\rightarrow(1,0,0)\rightarrow(1,1,0)\rightarrow(0,1,0)\rightarrow(0,0,0), 
\end{equation}
and  involves an energy exchange $U_\text{MH}$ between the system and the hot reservoirs.
The second cycle is
 \begin{equation}
 (0,0,0)\rightarrow(1,0,0)\rightarrow(1,0,1)\rightarrow(0,0,1)\rightarrow(0,0,0), 
\end{equation}
\end{subequations}
and involves an energy exchange $U_\text{MC}$ between the system and the cold reservoir. 
When the tunneling events in these sequences involve an electron that tunnels from reservoir L into the dot, and thereafter into reservoir R (or viceversa), it contributes to transport. The sum of all tunneling processes results in the generation of a finite electric current by the mere conversion of heat, only if a preferred direction is defined both by broken electron-hole and left-right symmetries~\cite{Sothmann15}.  We impose these conditions on the corresponding tunneling rates, requiring that the asymmetries,
\begin{equation}
\Lambda_\text{H}=\frac{\gamma_{\text{L}}(1,0)\gamma_\text{R}(0,0)-\gamma_\text{L}(0,0)\gamma_{\text{R}}(1,0)}{[\gamma_\text{L}(0,0)+\gamma_\text{R}(0,0)][\gamma_{\text{L}}(1,0)+\gamma_{\text{R}}(1,0)]}\ ,
\label{Eq:Lambda}
\end{equation}
and $\Lambda_\text{C}$ (obtained by replacing $\gamma_\alpha(1,0)\rightarrow\gamma_\alpha(0,1)$ in the previous expression) are finite. 

Solving the master equation~(\ref{Eq:master}), we get analytical expressions for the charge and heat currents. Despite these expressions being long and cumbersome (even for $V=0$),
we find a remarkably simple relation between them:
\begin{equation}
\label{IV0}
I_\text{R}(V{=}0)=e\left(\frac{\Lambda_\text{H}}{U_\text{MH}}J_\text{H}+\frac{\Lambda_\text{C}}{U_\text{MC}}J_\text{C}\right).
\end{equation}
It is instructive at this point to introduce $J_\text{in}$ and $J_\text{trans}$, where 
\begin{subequations}
\label{Eq:J_in&J_trans}
\begin{equation}
\label{Eq:heat_in}
J_\text{in} = J_\text{H}+J_\text{C}\ ,
\end{equation}
is the total heat flowing into dot M from the thermal reservoirs 
(reservoirs H and C), and 
\begin{equation}
\label{Eq:heat_trans}
J_\text{trans} = J_\text{H}-J_\text{C}\ .
\end{equation}
\end{subequations}
 is the heat that is transferred from reservoir H to C.  Note that a proportion of $J_{\rm trans}$ may {\it transit} through dot M on its way from reservoir H to C, even when none of it enters the electrical circuit (i.e. reservoirs L, R).  
Rewriting Eq.~(\ref{IV0}) in terms of  $J_{\rm in}$  and $J_{\rm trans}$,  one sees that $I_\text{R}(V{=}0)$ contains one part that depends on $J_{\rm in}$ as one might expect, but it also contains a term that depends on $J_{\rm trans}$, even though that energy does not go into the circuit. 

If the temperature and coupling of the hot and cold reservoirs are such that $J_{\rm in}=0$,
one still generates a charge current [which we here denote by $I^*=I(J_\text{in}=0)$] in the circuit equal to
\begin{equation}
\label{Inoheat}
I^*(V{=}0)=\frac{e}2\left(\frac{\Lambda_\text{H}}{U_\text{MH}}-\frac{\Lambda_\text{C}}{U_\text{MC}}\right)J_{\rm trans} \ .
\end{equation}
Whenever the charging energies  $U_\text{MH} $ and $U_\text{MC}$ are not equal,
 there is no symmetry reason for $I^*$ to be zero, so a typical system will exhibit  a charge current in the circuit
even though   there is no energy absorbed  from the thermal baths ($J_{\rm in}=0$).
In contrast, in the special case of $U_\text{MH}=U_\text{MC}$, one can show that 
$\Lambda_\text{H}=\Lambda_\text{C}$,
which means that $I^*(V{=}0)=0$.  To show this, one notes that if  $U_\text{MH}=U_\text{MC}$, then 
 the tunneling rates into dot M are not sensitive to which of the upper dot is occupied (H or C), so $\gamma_\alpha(1,0)=\gamma_\alpha(0,1)$, which implies $\Lambda_\text{H}=\Lambda_\text{C}$.  
It is thus essential that the two baths act differently on the system. We achieve this by taking different capacitive couplings.

\section{Results for finite power generation}
\label{Sect:finite-power}

\begin{figure}
\centerline{\includegraphics[width=0.7\columnwidth]{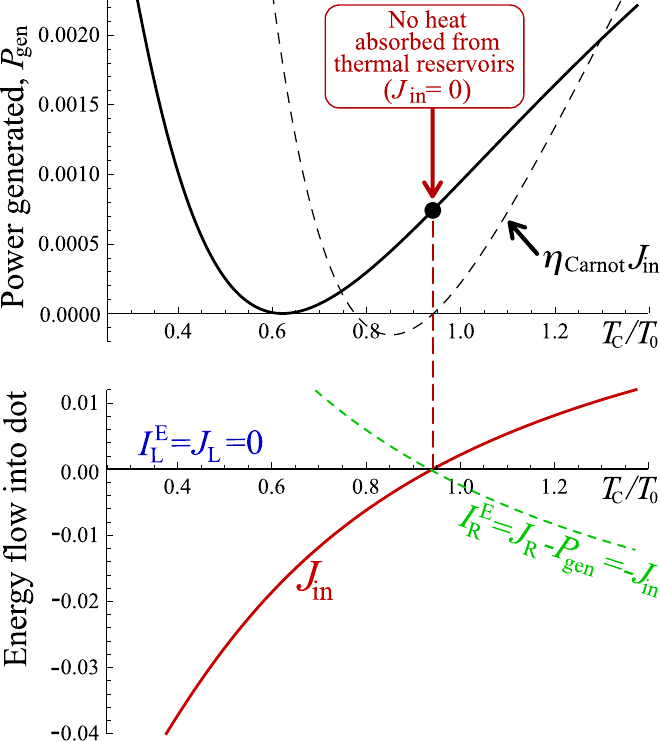}}
\caption{\label{Fig:basic-effect}
Upper plot: Power generated by the quantum-dot thermocouple (solid curve) as a function of $T_{\rm C}$ compared to the maximum power $\eta_{\rm Carnot}J_{\rm in}$ (dashed line) any conventional thermocouple could generate (see Sec.~\ref{Sect:thermocouple}). 
Lower plot: Energy currents into the system as a function of $T_{\rm C}/T_0$.
For both panels, the system parameters are given by the temperatures $\kB T_\text{H}/\gamma=16$, $\kB T_0/\gamma =8$, the dot energy levels $\eps_\text{M}=\eps_\text{H}=\eps_\text{C}=0$, the charging energies 
$U_\text{MH}/\gamma=3$,  $U_\text{MC}/\gamma=6$, $U_\text{HC}/\gamma=100$, and the coupling constants $\gamma_\text{H}/\gamma=1$, $\gamma_\text{C}/\gamma=0.5$, 
while $\gamma_\text{L}$ and $\gamma_\text{R}$ are given in Eq.~(\ref{Eq:gamma_L&R}).
}
\end{figure}

\begin{figure}
\centerline{\includegraphics[width=0.7\columnwidth]{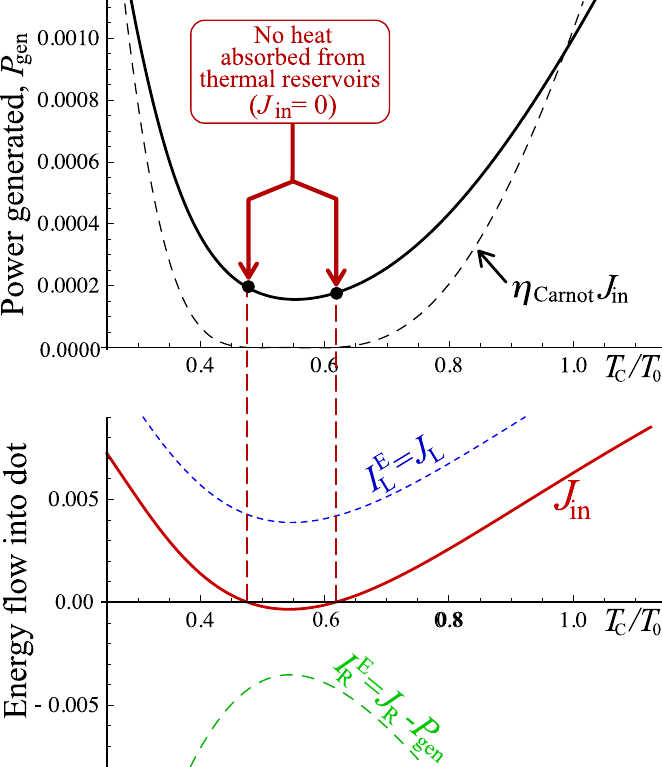}}
\caption{\label{Fig:basic-effect-other-parameters}
Upper plot:  Power generated by the quantum-dot thermocouple (solid curve) as a function of $T_{\rm C}$ compared to the maximum power $\eta_{\rm Carnot}J_{\rm in}$ (dashed line) any conventional thermocouple could generate (see Sec.~\ref{Sect:thermocouple}). 
Lower plot: Energy currents into the system as a function of $T_{\rm C}/T_0$.
For both panels, the system parameters are given by the temperatures $\kB T_\text{H}/\gamma =16$, $\kB T_0/\gamma =8$, the dot energy levels $\eps_{\rm M}=\eps_{\rm H}=\eps_{\rm C}=7\gamma$, the charging energies 
$U_\text{MH}/\gamma=5$,  $U_\text{MC}/\gamma=6$,    $U_\text{HC}/\gamma=3$, and the coupling constants $\gamma_\text{H}/\gamma=0.5$, $\gamma_\text{C}/\gamma=1$, 
while $\gamma_\text{L}$ and $\gamma_\text{R}$ are given in Eq.~(\ref{Eq:gamma_L&R}).
The bias $V=V_{\rm opt}(T_{\rm C})$ is the one that maximizes the power generated for each $T_{\rm C}/T_0$.
}
\end{figure}

\begin{figure}
\centerline{\includegraphics[width=0.7\columnwidth]{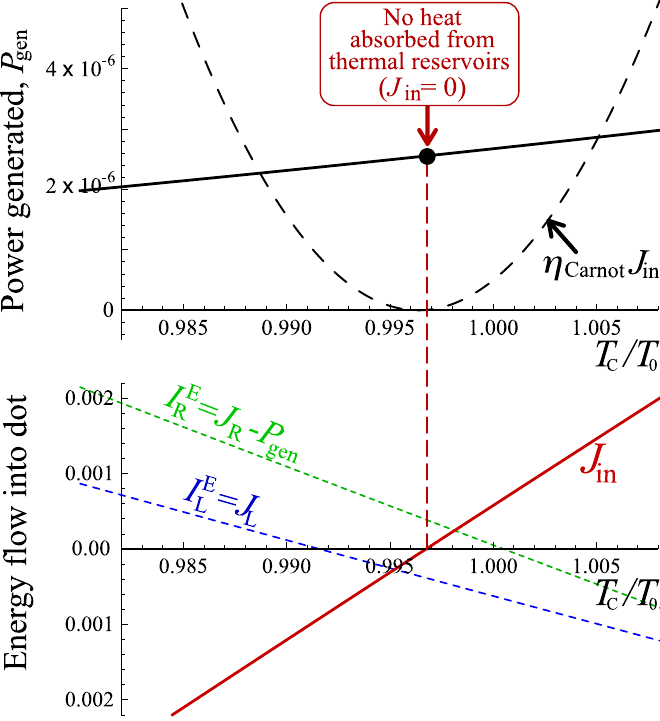}}
\caption{\label{Fig:basic-effect-poor}
Upper plot:  Power generated by the quantum-dot thermocouple (solid curve) as a function of $T_{\rm C}$ compared to the maximum power $\eta_{\rm Carnot}J_{\rm in}$ (dashed line) any conventional thermocouple could generate (see Sec.~\ref{Sect:thermocouple}). 
Lower plot: Energy currents into the system as a function of $T_{\rm C}/T_0$.
For both panels, the system parameters are given by the temperatures $\kB T_\text{H}/\gamma =16$, $\kB T_0/\gamma =8$, the dot energy levels $\eps_{\rm M}=\eps_{\rm H}=\eps_{\rm C}=6\gamma$, the charging energies 
$U_\text{MH}/\gamma=1$,  $U_\text{MC}/\gamma=12$,  and  $U_\text{HC}/\gamma=100$, and the coupling constants $\gamma_{\rm H}=\gamma_{\rm C}=\gamma$, 
while $\gamma_\text{L}$ and $\gamma_\text{R}$ are given in 
Eq.~(\ref{Eq:gamma_L&R-imperfect}). 
The bias $V=V_{\rm opt}(T_{\rm C})$ is the one that maximizes the power generated for each $T_{\rm C}/T_0$.
}
\end{figure}

For a finite power output,  the machine must drive a current 
against a finite potential difference $V$.  This occurs when the load's resistance is non-zero.
We also now consider the general case in which $U_{\rm HC}$ is finite, so the dot dynamics involves more cycles than just those in Eqs.~(\ref{Eq:cycles}), which can explore all eight possible states of the three-dot system.
Since in this general situation at any $V$, the  algebraic solution of the master equation, Eq.~(\ref{Eq:master}), and of the resulting charge and heat currents are extremely cumbersome, we refrain from presenting them here. Instead, in Figs.~\ref{Fig:basic-effect}-\ref{Fig:basic-effect-poor}, we plot the power output generated by our quantum-dot thermocouple as a function of the temperature of the cold reservoir, $T_\text{C}$, for various specific system parameters. For comparison, we also show the energy currents flowing from reservoir L and R into the dot M, $I^E_\text{L}$ and $I^E_\text{R}$, as well as the heat current flowing into the system from the hot and cold reservoirs, $J_\text{in}$. 

These plots show the possibility of generating finite power when the energy flow into the system from the hot and the cold reservoir vanishes.
Importantly, the effect is shown to be generic, since the variation of different system parameters does not alter the basic features. 
For the plots in this paper, we give all rates and energies in terms of a rate $\gamma$.
As a result the power generated $P_{\rm gen}$ and all energy currrents $J_\alpha$ 
are in units of $\gamma^2$ (we take Planck's contant $\hbar=1$). 
In all cases, we fix the temperatures of the hot reservoir and the load, $\kB T_\text{H}/\gamma=16$, $\kB T_0/\gamma=8$, and vary $\kB T_\text{C}$. In different plots we take different tunneling rates, $\{\gamma_\alpha\}$, 
but always take them to be smaller than the temperatures, which ensures the applicability of the master equation approach introduced in the previous section.  In each plot, for each value of $T_{\rm C}$, we find the bias voltage $V$ that maximizes the power generated.  We call this optimal bias $V_{\rm opt}(T_{\rm C})$.  We then plot the power generated  and the energy flows at $V=V_{\rm opt}(T_{\rm C})$ for each $T_{\rm C}$.

The plot in Figs.~\ref{Fig:basic-effect} and \ref{Fig:basic-effect-other-parameters} are for
\begin{subequations}
\label{Eq:gamma_L&R}
\begin{eqnarray}
\gamma_\text{L}(h,c)/\gamma&=&\delta_{h,0}\delta_{c,0} \ , \\
\gamma_\text{R}(h,c)/\gamma&=&1-\delta_{h,0}\delta_{c,0} \ .
\end{eqnarray}
\end{subequations}
This choice of $\gamma_\text{L}(h,c)$ and $\gamma_\text{R}(h,c)$, ensures that when dot H and C
are empty dot M is decoupled from reservoir R, but if either dot H or C is occupied
then dot M is decoupled from reservoir L.  
This means that the asymmetry parameters in Eq.~(\ref{Eq:Lambda}) are maximal, with $\Lambda_\text{H}=\Lambda_\text{C}=-1$.
All the other parameters are given in the figure captions.
As mentioned before, our model contains too many parameters to present a systematic study of the full parameter space, however the features we find appear to be generic. 
The following two paragraphs highlight this by discussing 
central points about Figs.~\ref{Fig:basic-effect} 
and \ref{Fig:basic-effect-other-parameters}.

In Fig.~\ref{Fig:basic-effect} we take a very large $U_\text{HC}$ (much larger than $\kB T_{\rm H}$), which means that the dots H and C are so strongly capacitively coupled that there is an extremely small probability of them being both occupied at the same time (similar to the limit $U_\text{HC}\rightarrow\infty$  taken in Sec.~\ref{Sect:V=0}).   
The power generated, $P_{\rm gen}$, vanishes quadratically at a certain value of $T_{\rm C}$.
This parabolic behaviour of $P_{\rm gen}= -I_{\rm R} V$, occurs because the current $I_{\rm R}$ 
changes sign (going linearly through zero), and
thus the bias generated also changes sign at this point ($V \propto I_{\rm R}$ for small $I_{\rm R}$). 
However, we clearly see that $P_{\rm gen}$ vanishes at smaller $T_{\rm C}$ than $J_{\rm in}$;
so $P_{\rm gen}$ is finite when $J_{\rm in}=0$ 
(marked by the filled circle in Fig.~\ref{Fig:basic-effect}).
The power generated at this point is purely an effect of the heat flow $J_{\rm trans}$, which passes from reservoir H to C without being absorbed by the thermocouple circuit.  We call this the {\it exotic} power generation. 
When $J_{\rm in}=0$, one has 
$I_\text{L}^E+I_\text{R}^E=0$, thus finite $P_{\rm gen}$ implies that there  is a net heat current out of 
reservoirs L and R, meaning that the total entropy of those reservoirs is dropping.  
Sec.~\ref{Sect:interpretation-nonlocal} explains why this is not a violation of the second law of thermodynamics.\footnote{Fig.~\ref{Fig:basic-effect} has  $I_\text{L}^E=0$, however
this has no physical consequences as $I_\text{L}^E$ is gauge-dependent (it depends on our choice of $E=0$). In contrast, the other quantities considered above, such as
$\big(I_\text{L}^E+I_\text{R}^E\big)$, $I_\alpha$ and $J_\alpha$, are gauge-independent and so have physical meaning.}
We do not show $J_{\rm trans}$ in Fig.~\ref{Fig:basic-effect}, as it is significantly larger than the other heat flows,
 it is a positive and monotonically decreasing function of  $T_{\rm C}/T_0$
 (it goes from 0.14 at $T_{\rm C}/T_0=0.2$ to 0.03 at  $T_{\rm C}/T_0=1.2$).

The plots in Fig.~\ref{Fig:basic-effect-other-parameters} show the same basic features as in Fig.~\ref{Fig:basic-effect}, despite the following differences in the choice of parameters. We no longer take $U_\text{HC}$ to be effectively infinite (i.e.~much larger than $\kB T_{\rm H}$). We take the energy-levels of the dot to be above the chemical potentials of the reservoirs. The coupling to reservoirs L and R are still given by  Eq.~(\ref{Eq:gamma_L&R}), ensuring the optimal asymmetry for the thermoelectric effect.
The decreased capacitive coupling $U_\text{HC}$ (of the order of temperature) allows all the states of the three dots to play an active role in the dynamics (unlike in the limiting case  $U_\text{HC}\rightarrow\infty$
considered in Sec.~\ref{Sect:V=0}). This however appears to have little qualitative effect on the physics (one can even take $U_\text{HC}=0$ without considerably changing the features that interest us).
However, the shift of the energy levels with respect to the electrochemical potentials is directly responsible for the only qualitative difference between the results in Figs.~\ref{Fig:basic-effect-other-parameters} 
and \ref{Fig:basic-effect}, namely that $J_{\rm in}$ now vanishes at two points in Fig.~\ref{Fig:basic-effect-other-parameters}.  This can be explained as follows.
When $T_{\rm C}$ is large, then $J_{\rm in}$ is positive,
but as one reduces  $T_{\rm C}$, the heat flow $J_{\rm in}$ goes to zero and becomes negative.
However, as one reduces  $T_{\rm C}$ even further $\kB T_{\rm C}$ becomes significantly less 
than $\eps_{\rm C}$, and so the dynamics of dot C start to freeze out  (thermal fluctuations of dot C become extremely rare, as $\epsilon_\text{C}-\mu_\text{C}\gg\kB T_\text{C}$). 
Thus, when $T_{\rm C}$ is sufficiently small, reservoir C becomes effectively decoupled from the rest of the system, the only heat flow is that from reservoir H to reservoirs L and R, and $J_{\rm in}$ becomes positive again.
The power generated remains positive at all $T_{\rm C}$ because of the exotic contribution 
coming from $J_{\rm trans}$. Another consequence of the off-resonant level positions chosen here, is that both the energy current out of the left and the right reservoir are in general non-zero. As required from energy conservation, they are however opposite, $I^E_\text{L}=-I^E_\text{R}$, at the point of vanishing  $J_{\rm in}$.
Again $J_{\rm trans}$ is positive and significantly larger than the other heat flows for all $T_{\rm C}/T_0$,  
as we show  
in the inset of Fig.~\ref{Fig:entropy}.  It drops for small $T_{\rm C}$, because of the above mentioned freezing of the dynamics of  dot C.

The plots in Fig.~\ref{Fig:basic-effect-poor} are for an imperfect thermocouple, where dot M is coupled to both reservoir L and reservoir R for all occupations of the of upper dots 
($h$ and $c$).  We replace Eq.~(\ref{Eq:gamma_L&R}) with  
\begin{subequations}
\label{Eq:gamma_L&R-imperfect}
\begin{eqnarray}
\gamma_\text{L}(h,c)/\gamma &=& (2 +\delta_{h,0}\delta_{c,0})/3, \\ 
\gamma_\text{R}(h,c)/\gamma &=& (3-\delta_{h,0}\delta_{c,0})/3.
\end{eqnarray}  
\end{subequations}
This means that the coupling strength is only varying by $1/3$ as $h$ and $c$ change and 
the asymmetry parameters in Eq.~(\ref{Eq:Lambda}) are $\Lambda_\text{C}=\Lambda_\text{H}=-0.2$.
The other parameters chosen for this plot are given in the figure caption.
The figure shows that the exotic power generation still exists for this
imperfect pump, although its magnitude is greatly reduced (just as the traditional power generation is greatly reduced).

\section{Interpretation}
\label{Sect:interpretation}

\subsection{Coulomb heat drag}
\label{Sect:interpretation-drag}

The charge current we found in Sec.~\ref{Sect:V=0} for $J_{\rm in}=0$
 is due to a form of Coulomb heat drag, in which 
the heat flow from reservoir H to reservoir C, 
{\it drags} a heat and charge currents from reservoir L to R. 
This is possible due to the capacitive coupling between the quantum dots. 
However, in other drag effects \cite{Sanchez10,Sanchez11},
there is a transfer of energy associated with one current dragging the other, which corresponds to finite $J_{\rm in}$.
Hence, this drag effect without energy transfer is extremely unusual.

This heat drag effect can occur for any $T_{\rm C}  < T_{\rm H}$ irrespective of the value of $T_{\rm C}/T_0$, since it only requires a heat flow from reservoir H to reservoir C. The quantum dot thermocouple then produces more work than a classical one (see the discussion in Sec.7). Remarkably, this does not only occur in the configurations where $J_{\rm in}\approx 0$, but also in a relatively wide parameter range around this points, including the case $T_{C}  \gtrsim T_0$.


\subsection{Non-locality and entropy production}
\label{Sect:interpretation-nonlocal}

\begin{figure}
\centerline{\includegraphics[width=0.95\columnwidth]{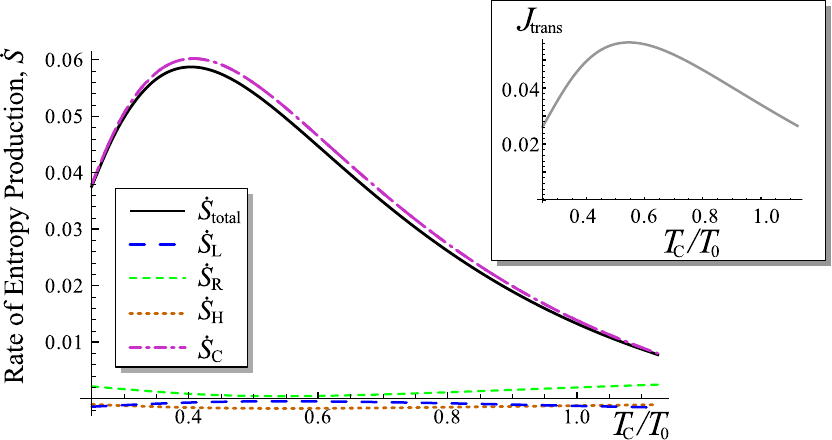}}
\caption{\label{Fig:entropy}
The entropy generated in each reservoir (main plot) and $J_{\rm trans}$ (inset) for the system in
Fig.~\ref{Fig:basic-effect-other-parameters}.  The entropy generated in reservoirs L, R and H are small 
in magnitude and their sum is negative, however the entropy generated in reservoir C is large and positive.  
Thus, the second law of thermodynamics is not violated.
We see that the rate of entropy production broadly follows $J_{\rm trans}$ (inset).
We do not show plots of entropy production and $J_{\rm trans}$ for the system in 
Fig.~\ref{Fig:basic-effect}, as the main features are generic to all of them.
}
\end{figure} 

Despite the unusualness of the Coulomb drag without energy transfer discussed above,
the situation is not paradoxical as long as $V=0$. The case  $V=0$ corresponds to a load with zero resistance, so nothing opposes the charge flow,
and so no work is required to generate a finite charge current.
In contrast, as soon as the  load's resistance is finite, the charge current $I$ must flow against a finite voltage bias, $V$, then work must be performed on the load.
It would be natural to assume that this work must be provided by the thermal reservoirs, i.e. that  the charge current $I$ can only flow if $J_{\rm in}\neq 0$.
However, something completely different can be observed in the results in Sec.~\ref{Sect:finite-power}:
there we show that finite power can be generated even when $J_{\rm in}=0$.
We call this {\it exotic} power generation. 
We observe that the power generated is equal to the heat absorbed by dot M 
from reservoirs L and R, as there is no heat flow into dot M from anywhere else when $J_{\rm in}=0$.
Thus the dot is sucking heat from these reservoirs (which are both at the same temperature), to turn it into useful work.
This obeys the first law of thermodynamics (energy conservation),
but -- if we ignore the presence of reservoirs H and C -- it appears to violate the second law, since it reduces the entropy of reservoirs L and R.

However, if we now turn to reservoirs H and C, we see that there is a heat flow from hot to cold,
and this increases the total entropy involved.  
An elegant way to verify this is the prove in Ref.~\cite{vandenBrueck-review} that any such master equation respects the laws of thermodynamics, thus we know that the four reservoirs and three dots
together do produce entropy.  
What is more, one can rewrite the total rate of entropy production,
$\big(\rmd S_{\rm pr}\big/\rmd t\big)= \sum_\alpha \big(\rmd S_\alpha\big/\rmd t\big)$, as
\begin{equation}
{\rmd S_{\rm pr} \over \rmd t}= {J_{\rm in} \over 2}\left({2 \over T_0}- {1 \over T_{\rm H}} - {1 \over T_{\rm C}} \right)
+ {J_{\rm trans} \over 2} \left( {1 \over T_{\rm C}} - {1 \over T_{\rm H}} \right) - {P_{\rm gen} \over T_0}. \ 
\end{equation}
Now, since we know that $\big(\rmd S_{\rm pr}\big/\rmd t\big) \geq 0$ for our system, see also Fig.~\ref{Fig:entropy}, one knows that
\begin{eqnarray}
P_{\rm gen} \leq {J_{\rm in} \over 2}\left( 2-{T_0 \over T_H} - {T_0 \over T_C} \right)
+ {J_{\rm trans} \over 2} \left( {T_0 \over T_C} - {T_0 \over T_H} \right) .
\end{eqnarray}
Hence, the second law allows the power generated, $P_{\rm gen}$, to be finite and positive, 
even if $J_{\rm in}=0$, as long as $J_{\rm trans}$ is finite and positive.
The entropy increase in reservoir C compensates for the reduction of entropy in the other reservoirs (see  Fig.~\ref{Fig:entropy}).
Thus the apparent paradox is resolved by the fact the power generation has a non-local nature.  The entropy generation caused by the heat flow between one pair of reservoirs (H and C) enables the conversion of heat into work at another pair of reservoirs (L and R), even though
there is no heat (or energy) flow between these two pairs of reservoirs and even though 
they can be arbitrarily far apart.

A ``poetic'' 
way of expressing this non-locality 
is to say that there is a spatial separation of first and second laws of thermodynamics.
The thermocouple manifests the first law by converting heat in reservoirs L and R into work.  
At the same time, some distance away, entropy is being generated by the
flow of heat from reservoirs H and C, manifesting the second law.
While we have shown that this behaviour is not paradoxical, it certainly is exotic!

\subsection{Towards experimental observation}
\label{Sect:expt}

Transport through capacitively coupled quantum dots in multiterminal conductors has been 
measured~\cite{Chan02,Hubel07,McClure07,Thierschmann15a}. 
The effect of large charging energies on the thermoelectric response of small tunnel junctions~\cite{Amman91} and quantum dots~\cite{Beenakker92} is well-known experimentally~\cite{Staring93,Dzurak93}.
Our proposed device is close to those very recently realized in 
experiments~\cite{Roche15,Hartmann15,Thierschmann15b}, 
in which a quantum dot generates a charge current between two reservoirs (L and R)
by the rectification of thermal fluctuations in a third capacitatively-coupled reservoir 
(the heat source).  In those systems, it is likely that the dot with
tunnel-coupling to reservoirs L and R also has at least a weak capacitative coupling to thermal fluctuations in its cold environment (including reservoirs L and R).  Thus, these experimental systems 
 have a strong similarity with the four-terminal set-up that we discuss. 

However, to unambiguously show the exotic power generation experimentally, it would be desirable to have a 
more controlled coupling to the cold environment.  This could be done by capacitatively coupling a third
dot to the systems in Refs.~\cite{Roche15,Hartmann15,Thierschmann15b}, which we believe is achievable
with the technologies used in those works. 
More challenging will be to show  experimentally that  $J_{\rm in}=0$, for this it will be crucial to measure the heat currents accurately.
For this, the capacitive coupling to the heat source and sink permits the different charge fluctuations to be resolved experimentally~\cite{Fujisawa06,Kueng12}. Therefore, $J_\text{H}$ and $J_\text{C}$ can be measured~\cite{Sanchez12}, so it would be possible to verify that a finite charge current $I$ flows when the condition $J_\text{in}=0$ is met. It is already possible to manipulate locally the temperature of the reservoirs~\cite{Hartmann15,Thierschmann15b,Roche15,Thierschmann15a}, thus getting a knob for controlling the heat currents and ensuring $J_\text{in}=0$.

\begin{figure*}
\centerline{\includegraphics[width=0.9\textwidth]{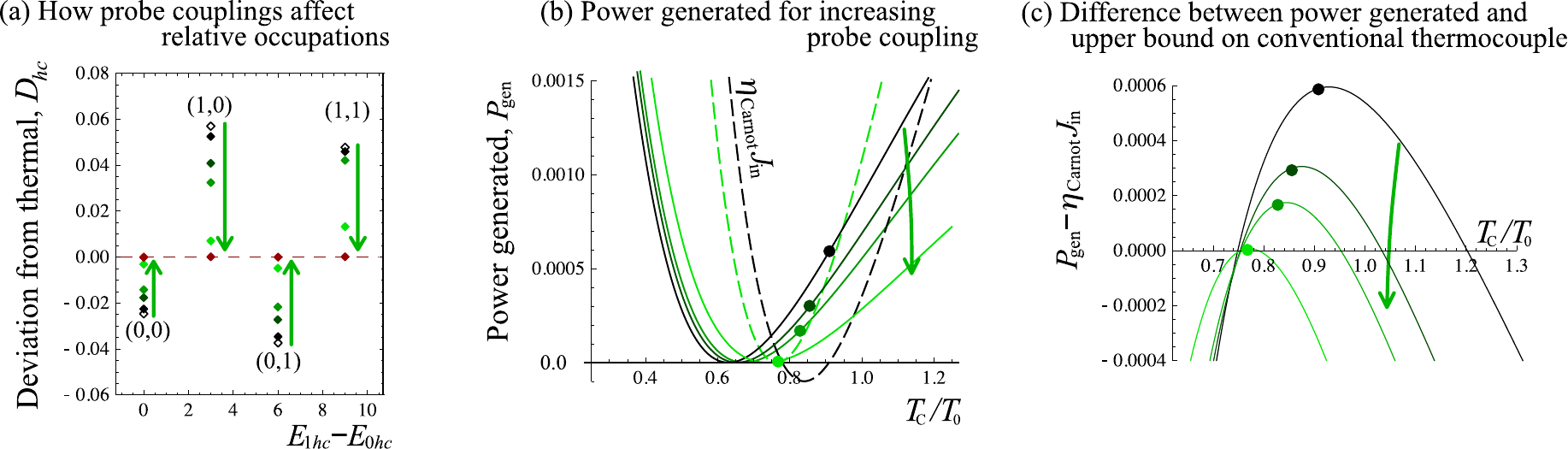}}
\caption{\label{Fig:probe-effect}
The same system parameters as in Fig.~\ref{Fig:basic-effect}, but with finite coupling to the temperature probe.
(a) How dot-M's state becomes increasingly thermal as the probe coupling is increased.
We show a scatter plot of $D_{hc}$ --- as defined in Eq.~(\ref{Eq:D_hc}) ---
versus $(E_{1hc}-E_{0hc})$ with $(h,c) =(0,0),(1,0),(0,1)$ and $(1,1)$.
As the coupling to the probe $\gamma_\text{P}$ is increased, 
the scatter of the points reduces towards zero.
Here we show  
$\gamma_{\rm P}/\gamma = 0.1,0.5,1,10$  (filled diamonds going from black to green),
with the arrows indicating increasing probe coupling.
We also show the two limits; $\gamma_{\rm P}  \to 0$ 
(open black diamond, for this we actually take  $\gamma_{\rm P}=0.001\gamma$)
and $\gamma_{\rm probe} \to \infty$ 
(filled red diamond, for this we actually take $\gamma_{\rm P}=1000\gamma$).
(b) The power output as a function of $T_\text{C}$ for different probe couplings (solid curves) for 
$\gamma_{\rm P}/\gamma = 0.1,0.5,1,10$. 
The filled circles on each curve indicate the value of $P_{\rm gen}$ when $J_{\rm in}=0$,
we see that this goes towards zero as the coupling to the probe increases.
The upper bound on conventional thermocouples, $\eta_{\rm Carnot}J_{\rm in}$, 
are indicated by the dashed curves for weak and strong probe coupling 
(black for $\gamma_{\rm P}  = 0.1\gamma$ and green for $\gamma_{\rm P}  = 10\gamma$).
Note that
for strong probe coupling the dot power output is below the bound for a conventional thermocouple for all $T_{\rm C}$.
This is seen more clearly in (c), which plots the difference in power generation between our quantum dot and the upper bound on a conventional thermocouple (again for $\gamma_{\rm P}/\gamma= 0.1,0.5,1,10$).
}
\end{figure*} 

\section{Thermalization induced by a probe reservoir}
\label{Sect:thermalize}

Our quantum dot model differs from that of a conventional thermocouple in a number of ways; 
such as the fact that dot M has discrete states and that it is in the Coulomb blockade regime.
Yet, we claim that the exotic power generation is the consequence of the  
non-thermal nature of the distribution in dot M, 
rather than one of the other properties of the dot.  

To verify this, we model the thermalization processes within dot M phenomenologically, 
by considering an additional probe reservoir (P),  which is tunnel-coupled to dot M,
as indicated in Fig.~\ref{Fig:system}.
The working principle of the probe \cite{MBSJ-local-temperature-2014,BS-local-temperature-2014,Stafford-local-temperature2014} is analogous to the one of a voltage probe~\cite{Buttiker88}, with the additional feature of probing the temperature. In other words, the probe is treated as an electronic reservoir in thermal equilibrium 
 at a temperature $T_\text{P}$ and  a chemical potential $\mu_\text{P}$, which have to be self-consistently determined by requiring that both the charge and heat currents, Eq.~(\ref{Eq:currents}), into the probe reservoir are equal to zero, $I_\text{P}=J_\text{P}=0$.
 
In order to evaluate the currents into the probe and to extract  $T_\text{P}$ and $\mu_\text{P}$ that make them vanish, we treat the probe reservoir on the same footing as the other electronic reservoirs introduced in Sec.~\ref{Sect:quantum-dot}. 
The coupling between dot M and the probe P is
\begin{eqnarray}
\Big[ \Gamma_\text{P} \Big]_{\, mhc}^{\, m'h'c'}\!\!\!\!\!\!
&=&\!\!\! \delta_{m,1-m'}\delta_{h,h'}\delta_{c,c'}\ \gamma_\text{P}\ 
f_\text{P}\left(\Delta^{m'h'c'}_{mhc} \right) \ ,
\end{eqnarray}
with Eq.~(\ref{Eq:M-in-terms-of-Gammas})'s sum over $\alpha$ 
extended to include P.
Then, $I_\text{P}$ and $J_\text{P}$ are given by 
Eqs.~(\ref{Eq:currents}) with $\alpha=\text{P}$.

If the probe is much more weakly coupled to dot M than the other two electronic reservoirs, ($\gamma_\text{P}\ll\gamma_\text{L,R}$), it does not influence the dynamics of the 3-dots system, i.e. it does not alter the results discussed in Sec.~\ref{Sect:V=0} and~\ref{Sect:finite-power}.  In contrast, if $\gamma_\text{P}$ is comparable with $\gamma_\text{L,R}$, or even larger, then the probe mimics phenomenologically the process of thermalizing transitions within dot M
(i.e. transitions that take the dot towards thermal equilibrium, without changing its average energy or charge).
In other words, the state of dot M becomes 
closer to thermal as we increase the coupling to the probe reservoir.

To measure
how far dot M is from a thermal state, we first note that if it were
in a thermal state corresponding to equilibrium with 
the probe reservoir, then fluctuations
 of $m$ induced by the probe reservoir in both directions 
($1 \to 0$ and $0 \to 1$) would have the same rate, irrespective of the state of the other  dots.
In other words, dot M is in a thermal state in equilibrium with the probe reservoir P if
$\big[\Gamma_{\rm P}\big]_{0hc}^{1hc}p_{1hc} 
= \big[\Gamma_{\rm P}\big]_{1hc}^{0hc}p_{0hc}$ for all $h,c$.
A little algrebra shows that this thermal state corresponds to 
\begin{eqnarray}
\frac{p_{1hc}}{p_{0hc}} 
\,=\, \exp\left[-{E_{1hc}-E_{0hc}-\mu_{\rm P} \over \kB T_{\rm P} }\right] \ ,
\end{eqnarray}
for all $h,c$.
Thus, we define the following measure of how far dot M is from thermal equilibrium,
\begin{eqnarray}
D_{hc} &=& \frac{p_{1hc}}{ p_{0hc}}  
\exp\left[{E_{1hc}-E_{0hc}-\mu_{\rm P} \over \kB T_{\rm P}}\right]  -1 \ .
\label{Eq:D_hc}
\end{eqnarray}
Fig.~\ref{Fig:probe-effect}(a) shows how increasing the coupling $\gamma_\text{P}$ to the  probe
tends to reduce $D_{hc}$ for all $h,c$, indicating that the steady state of dot M becomes closer to that of the thermal state with temperature $T_{\rm P}$.
Fig.~\ref{Fig:probe-effect}(b) shows that as dot M's state becomes closer to the thermal state, 
the power generated at $J_\text{in}=0$ goes to zero.  Thus the exotic power generation that we discuss in this article is a consequence of the fact that dot M is maintained in a non-thermal state by its couplings to 
reservoirs H and C.

\section{Comparing with a conventional thermocouple}
\label{Sect:thermocouple}

A conventional thermocouple corresponds to the system in Fig.~\ref{Fig:system}(a), with the quantum dot replaced by a macroscopic reservoir M that is larger than the electron's thermalization length, so  it equilibrates at a temperature $T_{\rm M}$ between $T_\text{H}$ and $T_\text{C}$.  
The role of the coupling to reservoirs H and C is to establish the temperature $T_{\rm M}$.  
In the steady-state, $T_{\rm M}$ is the temperature which ensures that the heat-flow out of reservoir M into the thermoelectrics is equal to the heat flow into reservoir M from reservoirs H and C, 
which we call $J_{\rm in}$.  Given $J_{\rm in}$ and $T_{\rm M}$, the thermocouple's efficiency, 
$\eta=P_{\rm gen}\big/J_{\rm in}$, cannot exceed 
Carnot efficiency $\eta_{\rm Carnot}= (1-T_0/T_{\rm M})$.  
Thus
\begin{eqnarray}
P_{\rm gen} \ \leq \ \eta_{\rm Carnot} \,J_{\rm in} \ ,
\label{Eq:naive}
\end{eqnarray}
with the equals sign holding only if the two thermoelectrics are Carnot efficient,
corresponding to them both having a figure of merit $ZT \to \infty$.

We can immediately see that a system like ours,
which generates finite $P_{\rm gen}$ 
when $J_{\rm in}=0$, violates the bound in Eq.~(\ref{Eq:naive}).  
Here, we wish to see over what range of parameters our
system violates this bound.  
To do this, we need to associate a temperature, $T_{\rm M}$, 
to the non-thermal state of the quantum dot, so we can calculate $\eta_{\rm Carnot}$.
There is no unique way of defining such a temperature, however a natural way to do so is to say it is the temperature that a probe would measure if placed in thermal contact with the dot
\cite{MBSJ-local-temperature-2014,BS-local-temperature-2014,Stafford-local-temperature2014}.
As such, we can use
probe reservoir P considered in the previous section in the limit of very weak coupling to the dot ($\gamma_{\rm P}\ll \gamma_{\rm L,R}$), so it has negligible effect on dot-M's state. Although dot M is in general in a non-thermal state, we can associate to it an {\it effective temperature} $T_{\rm M}$, by defining $T_{\rm M}$  as the temperature of the probe reservoir when there is no charge or heat current flow between dot M and reservoir P,   i.e. $T_{\rm M}\equiv T_{\rm P}|_{I_{\rm P}=J_{\rm P}=0}$. Note that Ref.~\cite{Stafford-local-temperature2014} argued that for having a meaningful definition of $T_\text{M}$, it is crucial that the probe's coupling to the dot is the same at all energies, so the probe is sensitive to the whole energy distribution of the dot.

Having defined an effective temperature for dot M, and knowing the heat-flowing into it, $J_{\rm in}$, we can plot the upper bound given by Eq.~(\ref{Eq:naive}) for any given parameters of the model. This is represented as a dashed curve in all plots, see Figs.~\ref{Fig:basic-effect}-\ref{Fig:probe-effect}.  
We see that the power generated by our machine (solid curves) exceeds this classical bound over a broad range of parameters,  including when
$T_{\rm C}$  is equal to or above $T_0$.

Note that $\eta_{\rm Carnot}J_{\rm in}$ will be negative whenever $J_{\rm in}$ and $T_{\rm M} -T_{\rm 0}$ are of opposite sign.  For example,  in Fig.~\ref{Fig:basic-effect} one has $J_{\rm in}<0$ for $T_{\rm C}/T_0 < 0.94$,
while $T_{\rm M} < T_{\rm 0}$ only for $T_{\rm C}/T_0 < 0.77$.  Thus for  $0.77 < T_{\rm C}/T_0 < 0.94$, 
the dot M is hotter than reservoirs L and R but, since $J_{\rm in}<0$,  we know that energy is flowing from reservoirs L and R into the dot. In a macroscopic thermocouple, this can occur only if one provides energy to the system by replacing the load with a power supply that forces the current flowing through the circuit. In this case, power is not generated, but absorbed by the system, i.e. $P_{\rm gen} <0$.   
In contrast, our quantum dot system generates power, $P_{\rm gen} >0$, even in this range of parameters, for the
reasons discussed in Sec.~\ref{Sect:interpretation}.

Figs.~\ref{Fig:probe-effect}(b) and (c) show that as dot M's state becomes closer to that thermal state, the violation of Eq.~(\ref{Eq:naive}) goes away, leaving a system with $P_{\rm gen}= \eta J_{\rm in}$, 
where $\eta$ is the system's efficiency, and it is less than the Carnot efficiency $\eta_{\rm Carnot}$. For example,  for $\gamma_{\rm P}=10\gamma$ in Fig.~\ref{Fig:probe-effect}(b),  $\eta$ is of order $0.25\eta_{\rm Carnot}$. 
This confirms our conclusion that a non-thermal steady-state distribution of electrons in dot M 
is necessary for it to generate more power than any conventional thermoelectric could.


\section{Conclusions}

We have investigated the unexpected properties that arise in thermoelectric conductors when they become smaller than the lengthscale in which electrons thermalise. For this, we have considered a quantum dot based thermocouple 
in a four terminal geometry: two terminals support the electrical current, with the other two being the hot and cold environments with which energy is exchanged. Our system can supply electrical power to a load. Surprisingly, this can be done also in configurations in which no net heat is absorbed from the thermal reservoirs.
This exotic power generation can be explained in terms of non-local Coulomb heat drag in which the heat flow from the hot to the cold reservoir induces both a heat and a charge flow in the capacitively coupled electronic circuit. 
This effect should be realizable in systems very close to those in recent 
experiments~\cite{Roche15,Hartmann15,Thierschmann15b}, as outlined in Sec.~\ref{Sect:expt}.

We argue that this exotic power generation relies on the fact that quantum dot M is maintained in an non-thermal state, by showing that it disappears when 
relaxation effects destroy the non-thermal state. 
As such it cannot occur in a conventional macroscopic thermocouple heat-engine.
We compare our quantum dot thermocouple to its macroscopic equivalent, and find that avoiding the thermalization of carriers can help to improve the thermoelectric performance. In particular, the generated power can be larger than that of any macroscopic thermocouple, even when the latter works with Carnot efficiency. 
In the long term, we hope that this observation will be applied in thermoelectric and photovoltaic applications, since in both cases a large part of the heat flowing into the central region of
the thermocouple (or the central region of the p-n junction in photovoltaics) flows out again into
the cold environment, without contributing to the power generation.  
It would be a great benefit if a part of this ``lost'' energy flow could be used to make
electrical power in the manner presented here.

The same effect could be achieved with non-interacting conductors in magnetic fields,
via a Nernst effect~\cite{Stark14,Sothmann14}. 
Note that Refs.~\cite{Stark14,Sothmann14} assumed that the reservoirs that carry the charge current (reservoirs L and R in our Fig.~\ref{Fig:system})
are at temperatures chosen such that there is no heat flow out of them. In this situation, energy conservation guarantees that $P_{\rm gen}=0$ for $J_{\rm in}=0$.  However, for other temperatures, one expects to observe the exotic power generation that is discussed in the present paper.

An open question is how to engineer the exotic power generation when 
the capacitative coupling is replaced by exchange of photons~\cite{Bergenfeldt14} or phonons \cite{Rutten09,Entin10,Ruokola12,Jiang2012}.  
Not only is this crucial for thermoelectric and photovoltaic applications, 
it would help to elucidate the fundamental requirements for the effect.

As a step in this direction, it may be worthwhile studying the linear-response regime in more detail.
The results presented here are valid for arbitrary $T_\text{H}$, $T_\text{C}$ and $T_0$; while the observed effect is most prominent for very different temperatures, it also exists when the temperature differences are small. In this limit, one could analyze the effect using a multi-terminal Onsager treatment~\cite{Jiang2014b,jacq-whit-meir-buttiker2012}, which is more universal, but only works when there is a linear relationship between thermodynamic forces and currents.

\section*{Acknowledgements}

R.~W. and R.~S. thank D.~Weinmann and B.~Sothmann, respectively, for helpful discussions.
Part of this work was carried out thanks to the COST Action MP1209 ``Thermodynamics in the quantum regime'' (STSM visit to the Institute for Quantum Information, RWTH Aachen, and the ``Second Conference on Quantum Thermodynamics'', Mallorca, 2015).
We acknowledge financial support from the CNRS PEPS funding \hbox{``PERCEVAL''} (R.~W.), the Spanish MICINN Juan de la Cierva program and MAT2014-58241-P. (R.~S.),  the Knut and Alice Wallenberg Foundation and the Swedish VR (J.~S.), and the Alexander von Humboldt Foundation (F.~H.).



\end{document}